\documentstyle[12pt,aasms4]{article}
\begin{document}
\begin{center}

{\Large \bf  On the ionizing continuum in active galactic nuclei:
 clues from ISO} 

\vskip 0.4in
{\bf M. Almudena Prieto$^{1,2}$ and Sueli M. Viegas$^3$}

{ $^1$Instituto de Astrof\'\i sica de Canarias, Spain}\\
{ $^2$Max-Plank-Institut fuer extraterrestriche Physik, Garching, Germany}\\
{ $^3$Instituto Astron\^omico e Geof\'\i sico, S\~ao Paulo, Brazil} \\

\end{center}

\vskip 0.4in
\begin{abstract}
The ISO coronal line spectrum of the brightest Seyfert 
galaxies from the CfA
sample is presented and modeled. ISO observations of
[O IV] 25.9 $\mu$, [Ne V] 14.3 $\mu$,
[Mg VIII] 3.02 $\mu$ and [Si IX] 2.58 $\mu$ lines are presented; 
 their relationship with 
the  soft part of the ionizing spectrum from 50 to 300 eV is 
investigated.  Pure photoionization
models reproduce the line ratios, setting ranges for the ionization
parameter and the optical depth of the emitting clouds. On the basis
of the available data alone it is not possible to
 distinguish between a power-law or
a blackbody distribution as the intrinsic shape of the UV ionizing
spectrum.  However, for the brightest Seyferts analyzed, namely,
 NGC 1068, Circinus and NGC 4151, a black-body UV continuum is 
favored.

\end{abstract}

{\it Subject Headings:} galaxies: Seyfert - line: coronal - line:
formation - infrared: galaxies

\section{Introduction}

The near-to-mid IR spectrum is extremely rich in 
emission lines from many different ionization stages including molecular, 
nebular and coronal lines  
from different species. Coronal lines are particularly suitable for
deriving information about the UV to the soft X-ray
 region of the  ionizing spectrum as they
require photon energies above 50 eV.   Pure starbursts, where
[OIV] 25.9  $\mu$  is  generally not present or very week (Genzel et
al. 1998; Lutz et al. 1998)
 do require  photon energies below 50 eV. Thus, 
the study of coronal lines is unique for tracing 
the pure AGN power mechanism.
Important coronal lines, particularly from Fe,  are also present  in the
optical spectrum of 
some bright AGNs. However the larger extinction in the optical
and the fact that reliable detection of these lines requires medium to
high spectral resolution make the number of 
detections scarce. Alternatively, the IR region is less affected by
extinction and proves to be ideal for detecting a wide range 
of coronal species. The availability of ISO
allows us to make a systematic search for such coronal lines in
Seyferts galaxies.

The CfA sample of Seyfert galaxies    is the standard reference sample of
 active galactic nuclei 
in the nearby universe.  For some of the
brightest galaxies, a large wavelength range is covered, 
 in particular, the UV, optical, infrared (IRAS) and radio. 
However, due to the large extinction affecting the 
Seyfert II type, very poor information is derived 
from the UV region and few
reliable X-ray spectra are obtained for that class.
An alternative to get clues on that part of the electromagnetic 
spectrum is via the  study of the  IR region.
The sample of coronal lines studied in this work requires
 photon energies in 
the 50-300 eV range and therefore, their analysis can yield information 
about the extreme-UV-to-soft-excess continuum  in AGN.
With that aim in mind,
the brightest Seyferts from the CfA sample were selected for
observations with the SWS spectrometer on board ISO.
 Essentially, all those
sources that were detected with IRAS at 12 $\mu$ were selected. 
Among those, there
are well known galaxies (e.g. NGC 4151, NGC 1068, NGC 3727) 
that were object of
 individual observations within the ISO guaranteed 
time programs and were
not proposed for new observations with ISO.  ISO SWS spectra of the 
coronal lines
[O IV] 25.9 $\mu$, [Ne V] 14.3 $\mu$,
[Mg VIII] 3.02 $\mu$ and [Si IX] 2.58 $\mu$  were collected for all of
them.
The  sample is complemented 
with data from three brightest Seyfert 
 galaxies --- NGC 1068, Circinus (not CfA member) and NGC 4151--- 
for which SWS  spectra of those coronal lines are currently available in
the literature.
The final   sample comprises seven Seyfert 2, four Seyfert 1
 and one Seyfert 1.5 (Table 1).

\section{The ISO coronal spectra of Seyfert galaxies}

Our ISO observations were executed as follows.
For each source, four individual SWS line spectra with resolution of
about 200 km/s (SWS02 mode)  and  integration times per source between
1.5 and 2 hours were collected.  Although the selected sample includes
the brightest known Seyferts in the  IR, i.e.,  all having been
detected  at 12$\mu$ by IRAS,  that did not prove to be sufficient
when considering the sensitivity levels of the SWS02 mode. To keep
within reasonable exposure times, only the  coronal lines: [O IV] 25.9
$\mu$, [Ne V] 14.3 $\mu$, [Mg VIII] 3.02 $\mu$ and [Si IX] 2.58 $\mu$
were proposed for   ISO observations.   An additional  serendipitous
spectrum centered at about 4.6$\mu$ was also obtained.

Data reduction of the SWS spectra was done at the 
ISO spectrometer data center at MPE using the  Interactive Analysis
IA package. Special  routines built at MPE,
 aimed at improving spike removal, dark subtraction and 
flat fielding, were used.  A minimum of five scans  were  collected
per each ISO line spectrum to allow for deglitching. 

As each SWS spectrum combines the information of 12 individual detectors, 
each covering different  but overlapping regions of 
a given spectral window,
the extraction of the line spectrum requires 
a careful   process. 
In particular, as redundant information from 
various detectors is summed up at the center of the line spectrum,
  the signal-to-noise ratio 
 accumulated at the center of the spectral window  
is expected to be superior  than that at the edges
 where there is less overlapping information. 
That  translates into line spectra with higher  signal-to-noise at the
center but  progressively degrading towards the edges.
Nevertheless, the ISO line spectra  were 
centered at  the expected position of the corresponding line 
feature after
correcting  by the systemic velocity of the source. Thus,  
spectral features appearing at the expected position  might be 
considered of large  reliability. 

For each galaxy, the ISO line spectra are shown in
Figures 1a,b.
The spectral resolution of the SWS mode is between 300 km/s for the long 
wavelength region and about 150 km/s for the short one. Typically,   those
are the binning sizes used to combine the individual scans to produce
the final  line spectrum.

The derived line fluxes and errors are given in Table 1. The errors reflect
only the uncertainty in the continuum definition. Besides,
the absolute flux scale of the SWS is accurate to about 25\% (Schaeidt
et al. 1996). Also, uncertainties in the lines position of up  100 km/s
may  be expected  due to pointing errors (Feuchtgruber et al. 1997).

In all cases, the strongest lines detected are
[O IV] 25.9 $\mu$ and [Ne V] 14.3 $\mu$ lines whereas 
[Mg VIII] 3.02 $\mu$ and [Si IX] 2.58 $\mu$
are the most difficult  to measure  in both
Seyfert types. Due  to the low sensitivity level 
of the SWS in the short wavelength range, some of the
apparent features in the
ISO [Si IX] 2.58 $\mu$ and [Mg VIII] 3.02 $\mu$ spectra
 are uncertain. To keep on the safe side,
all the  reported fluxes for [Si IX] 2.58 $\mu$  and most for 
[MgVIII] 3.02 $\mu$ (Table 1)
 are given as upper limits or marked  uncertain. These upper limits
are either  the
integrated flux within the apparent feature or a 3 sigma
 limit (Table 1).

Mrk 817 is the only case with potential detection of both [Si IX] 2.58
$\mu$ and   [MgVIII] 3.02 $\mu$, as corresponding faint features
 at the  expected
position are seen  in the  ISO spectra
(Fig. 2). Additional  
possible detections of
[Mg VIII] 3.02 $\mu$  are in Mrk 266 and Mrk 533. We marked those as
uncertain since the apparent
 feature is  shifted by $\sim$200 km/s in both galaxies.
Besides,  there are the relatively stronger detections reported for 
Circinus, NGC 1068 and NGC 4151 
(Moorwood et al, 1996,   Lutz et 
al, 1997 and  Alexander et al. 1998, respectively),
where
 [Si IX] 2.58 $\mu$ and  
[Mg VIII] 3.02 $\mu$ are clearly seen. 

Regarding the [OIV]25.9 $\mu$ and [Ne V] 14.3 $\mu$ lines, they 
 appear at the corresponding systemic velocity within the 100 km/s
 range for
 most of the sources. There are three exceptions where the [OIV]25.9 $\mu$
line is found substantially blueshifted.  In  NGC 5548, [OIV]25.9
 $\mu$  is shifted by $\sim$-150 km/s, which is in fair agreement
 with the shift of  -134 km/s  measured by
 Penston et al (1984) in the [FeX] 6374A line.
In Mrk334 and NGC 5929,  [OIV]25.9 $\mu$ shows shifted by 
about -300 km/s. An equivalent shift is not apparent in  [Ne V] 14.3
 $\mu$. We note however
 that the general [OIV] and [NeV] line profiles in most of 
 the Seyfert analyzed shows
 more complex than a single Gaussian profile.
That prevents an accurate
determination of the central line peak and the corresponding line shifts,
 particularly if the S/N is not sufficiently high.

[FeII] 25.98 $\mu$ line was also found  in some of the galaxies in the
 sample. Due to its proximity to [OIV]  25.9 $\mu$, that line 
 can also be seen in the  ISO [OIV]25.9 $\mu$  spectrum.
 Detections of the FeII line are seen in Mrk 266, NGC 5929, Mrk 334 and
 Mrk 817 (Fig. 1ab).
 Due to the poor spatial resolution of ISO -- SWS aperture sizes range
 from 
 14x20 arcsec at  short wavelengths to   20x33 arcsec at long
 wavelengths -- the integrated
 [FeII]25.98 $\mu$  emission    may mostly 
 comes from   the star forming 
regions in the disc of these galaxies. As the present analysis focuses on the 
central AGN source, that line is not 
 considered for the  modeling purposes.
 
\section{Modeling the IR lines}

The presence of high ionization lines in the far infrared spectrum
of Seyfert 1 and 2 is probably the signature of high energy photons
produced in the central source, providing highly ionized gas. 
>From this point of view, the analysis of the 2.5 - 45  $\mu$ ISO spectrum of 
Circinus (Moorwood et al. 1996) indicated that the ionizing
radiation spectrum should have an UV bump, represented by a
 2$\times$ 10$^5$ K blackbody radiation, added to a
X-ray power-law, with spectral index $\alpha_X $ = -0.5.
 The IR lines are roughly reproduced, the
coronal lines being  fitted within a factor of 2.  A further
detailed analysis of the {\it multiwavelength emission-line and
continuum spectra} (Contini, Prieto \& Viegas 1998a),
indicates a more complex scenario for Circinus central region. That
 basically requires a weighted  contribution of
 clouds with different  densities and optical depth,
 some of them showing
the effect of high-velocity shocks.

The present analysis  is restricted to the set of coronal
lines  discussed in section 2, which
 were systematically observed by ISO for all the
galaxies in the sample. They 
are used  to test the ionizing continuum shape in these 
galaxies on the basis of  photoionization models only. 
More elaborated analysis based on a multiwavelength  analysis approach
-- as those  done for 
the Seyferts prototype  Circinus and NGC 5252
(Contini, Prieto \& Viegas 1998a,b) -- are out of the scope of this paper.

  The behavior of the observed line
ratios [Ne V]/[O IV], [Mg VIII]/[O IV] and  
[Si IX]/[O IV] versus the ionization potential (IP) of Ne$^{3+}$,
Mg$^{6+}$ and Si$^{7+}$ is shown in  Figure 2ab.
The brightest galaxies, for which clear detections of [Si IX] 2.58
$\mu$ and [Mg VIII] 3.02 $\mu$ 
exist, show a common  trend of decreasing line ratios with the IP.
The other galaxies of the sample shows also  a clear decrease toward  
[MgVIII]/[OIV]. However, if we extend the plot in order to include
[SiIX]/[OIV], the overall behavior is equally consistent with a 
decreasing trend or a valley shape -- [SiIX]/[OIV] larger than
[MgVIII]/[OIV] --.
If the detection of   [SiIX] 2.58 $\mu$  in Mrk 817 is confirmed, 
a valley shape is found.

Regarding    [Ne V]/[O IV] and [MgVIII]/[O IV] line ratios, 
the Seyfert 2 sample present values 
within the range  measured for Circinus and NGC 1068 (figure 2). A clear
 exception is NGC 5929, but this galaxy  presents a weak ISO coronal
line spectrum. In the case of the Seyfert 1 sample, the  [Ne V]/[O IV]
and  [Mg VIII]/[O IV] values are in all cases 
larger than in  NGC 4151. Although most of the   [Mg VIII] values 
are upper limits, note however that
the [NeV] and [OIV] lines are well determined for this sample. 

With the above caveats in mind,  
the present modeling  investigates the
 relationship between the  coronal line ratios and the
ionizing radiation spectrum required  to form the corresponding ions. 
The photoionization simulations are obtained with  the 
photoionization code Aangaba (Gruenwald \& Viegas 1992) 
which has been compared to
similar codes (Ferland et al. 1995). The basic assumptions 
considered in the current modeling are  plan-parallel symmetry,
 solar abundances, and a composite
ionizing spectra (blackbody plus an X-ray power law, or
two power-laws). For the X-ray component, we adopt
the canonical   power-law index  for AGN $\alpha_X$ = -0.7.
Firstly, we discuss the results for two types of UV continuum,
namely, a  2.$\times$ 10$^5$ K blackbody  and a  
power-law with power index $\alpha_{UV}$ = -2.0. In 
 both cases, the UV spectrum  reaches the X-ray regime 
 at E $\simeq$ 100 eV.
We  anticipate that  variations in the black-body temperature or the
UV spectral index do not introduce significant differences in the model
results; however,  these   are   more
sensitive to changes in the Ec value.

The critical density of the coronal emission lines is high, so the
results are practically independent of the gas density
as long as  $ne\leq$ 10$^5$ cm$^{-3}$. Thus, the two parameters that
are left free are  {\it the ionization parameter U }(ratio of
 the number density of ionizing photons and the gas density),
{\it and the cloud optical depth}, characterized by the optical
depth at the Lyman limit, $\tau_{ly}$.  

 The line ratios versus IP are shown in Figures 3a (UV blackbody) and
3b (UV power law). The top panel shows the results for
a given U and varying   $\tau_{ly}$, and the bottom panel the
theoretical ratios for an optically thick cloud 
($\tau_{ly}$ $\geq$ 30) at different U values.  Both set of models
produce [NeV]/[OIV] and [Mg VIII]/[O IV] values within the limits 
that are observed in the present sample. A first result illustrated by
the figures is that for
both set of UV ionizing continuum, an  overall 
``valley shape'' is  generally produced.

Blackbody
models with low $\tau_{ly}$ ($\leq$ 2) show a different behavior,
namely, the line ratios increasing with the ionization
potential (upper part in fig 3a). This is because the 
optically thin models tend to penalize the lower ionization lines, 
favoring the high ionization zone. The ratios are relative to
 [O IV]. This line originates in the O$^{3+}$ region which decreases
with $\tau_{ly}$
 faster than Mg$^{6+}$ and Si$^{7+}$ zones; thus,
[Mg VIII]/[O IV] and  [Si IX]/[O IV] tend to became higher 
 as $\tau_{ly}$ decreases. 

Conversely, blackbody models with low U value tend to smear out 
the valley shape 
 by producing slightly lower [Si IX]/[O IV] than  [Mg VIII]/[O IV] (fig 3a).
A similar effect  occurs as well  in the case of the UV
power law model, but  in this case the effect  is 
 much more dramatic as the [Si IX] and  [Mg VIII] lines tend to
vanish (lower part in fig 3b).

If  an UV power index  $\alpha_{UV}$ = -1.5 is  adopted, the results
are very similar, all the models showing a valley
shape even in the case of low  $\tau_{ly}$ values. In this case the
O$^{3+}$ zone is closer to the cloud edge facing the central 
radiation source and thus, less affected by matter-bounded
conditions. 
Regarding the UV black-body, variations in the temperature within a
few $ 10^5$ K do not
modify the general trend shown in Fig 3a.

Summarizing, within the set  of model parameters considered, both   UV models (blackbody  and  power-law) 
  predict  similar  results, namely, a 
valley shape relationship between the
coronal line ratios  and the ionization potential required to produce the
corresponding ions. This model result can  easily account for the case of
Mrk 817 if the [SiIX]2.58 $\mu$ and [Mg VIII] 3.02 $\mu$ lines turn to
be reliable detections; yet, it clearly departs from the most solid
 trend seen
in the brightest Seyferts, NGC 1068, Circinus and NGC 4151. In these cases,
the observed line ratios show a straight decreasing trend vs the
 ionization potential.

\subsection{Diagnostic diagrams}

 The four observed coronal lines allows us to construct 
two diagnostic diagrams. These are 
 discussed on the basis of
the    model predictions for the two
types of UV continuum considered.

Focusing first on the observed line ratios
(plotted   in Figures 4), 
a general trend   is suggested in  both  diagrams,  in the 
sense that both [SiIX]2.58 $\mu$ and [Mg VIII] 3.02 $\mu$ lines relative 
to [OIV]  tend to increase with
increasing [NeV]/[OIV]. This trend is surprising if considered that
most of the  [SiIX]2.58 $\mu$ and [Mg VIII] 3.02 $\mu$ measurements
are taken as upper limits.
 
 If considered the   UV black-body
models, the  observed trend   can be
reproduced   for
 U  $\leq$ U$_c = 0.1$,
and  optically thick clouds, $\tau_{ly} \sim>1$ (Fig 5a,b). 
The upper limits for
[MgVIII] and  [SiIX] will just move the points down in the diagram suggesting
lower U values. Indeed,  this is the case for the brightest Seyferts
which require U values below 0.01 (discussed in next section).

  In the case of  UV power law models (Fig. 6a,b), the predicted  line
ratios always  increase
with U, for  U $\leq$ U$_c$ = 1 (this being the main difference with
the black-body models).  Besides,  for
$\tau_{ly}$ $\geq$ 1.,  the theoretical ratios predicted by the power
law models become very similar.  The comparison with the
 observational data indicates similar range of U values as in the blackbody 
case but  the emitting clouds are closer to the optically thin case, with
values in the range   0.2 $\leq$ $\tau_{ly}$ $\leq$ 1.0 for most
of the objects.

\subsection{The brightest Seyferts}

In an attempt to account for the 
line ratio  trend shown in Fig.2 
by NGC 1068, Circinus and NGC 4151,  
the  UV black-body models  are explored
in more detail in this section. Models with 
an UV power law were also built but the results were less encouraging.

Varying the   black-body temperature  within a few of  10$^5$ K,
the best representation of  the
observed trend is  obtained for $ T\sim 2\times 10^5$ K and $\alpha_X$ = -0.7.
Furthermore, 
 the ionization parameter should remain low ( U $\simeq$ 0.03) and the
optical depth  be   larger than unity.  Larger U
or lower $\tau_{ly}$  tend to generate the ``valley shape''
discussed in section 3. In this case, the zone producing the
[O IV] line is not so wide, the [O IV] is weaker and the line ratios
become higher.

In all the models  discussed in former sections,
the adopted value for the energy break between the UV black-body 
and the
X-ray power-law was set to Ec= 100 eV. However, we may  expect
that slight variations  in Ec    significantly affect the line ratios,
particularly those involving SiIX and MgVII, as 100 eV is within the range of
IP values  required produce those ions. 

The final results are as
follows.  The best match to the data is found for Ec=120 eV,  U=0.01 and
optical depths about  1.5 or larger. Models with optical depth
between 1.5 and 30 produce very similar results provided  that the
ionization parameter remains low. The results are plotted in Fig. 7: 
only two set of models are shown.   The first set of models
corresponds to
 U=0.01 (two models are plotted for tau=1.5 and 30 respectively). 
The second set of
models corresponds to lower U=0.005 (plotted for tau=1.5 and 30). 
In
all cases, Ec=120 eV. For larger Ec,  the predicted [NeV]/[OIV] gets
too strong compared with the data (see also Fig 5ab). In fact, because the UV black-body continuum
is steeply decreasing at these energies, larger values of Ec would favor the 
production of O$^{+3}$ and Ne$^{+4}$ instead of increasing the higher
 ionized ions: i.e., 
 [NeV]/[OIV] tends to increase faster relative to  [Mg VIII]/[OIV] and  
[Si IX]/[OIV].

The  models  just described provide a rough approximation of the 
observed trend. No attempt was made to fit the individual  line
ratios because we expect that various clouds at different conditions
should contribute to the observed spectrum. In particular, the higher ionization
lines ([Mg VIII] and  [Si IX]) may be produced in clouds different
from those producing the lower ionization ones  ([O IV] and [Ne V]). 
In addition, 
changes in the relative abundances (current
modeling is for solar abundance) can play a major
role in fitting individual  line ratios.

 Recently, Binette (1998) discussed models for the coronal lines of Circinus 
and NGC 1068, assuming the ionizing continuum represented by two power laws,
with power index equal to 1.3 in the infrared-UV range and 0.7 in the X-ray 
range, and
joining at Ec = 2000 eV. Such a high Ec value
makes  the ionization structure of the gas
to be mainly determined by the UV continuum. His model includes radiative 
acceleration
of the coronal gas, which results in gas compression by factors 
of 1.6 to 2.6. 
From the fraction of unabsorbed ionized photons listed in his paper,
we deduce that
the optical depth of the cloud is also greater than unity. 
However, for both galaxies the ionization parameter is greater than 0.1.  
The difference with our models comes from the shape of the UV continuum
which in his case is flatter than that used in our models. This flatter spectrum coupled
to the increase of the gas density due to compression, makes the lower ionized
zone (where  [O IV], [Ne V], [Si VI], [Fe VII] originate ) narrower than the
higher ionized zone.  Thus  the trend of the line ratios decreasing with 
the ionization potential discussed above cannot be reproduced. For this,
lower values of U ( $\simeq$ 0.01) must be considered. But, in this case
the  Fe coronal lines relative to [Si IX] (as used by Binette)
 become too strong. 

 Binette's models and ours clearly   indicate that a good fit to the
coronal lines can only be achieved if the contribution from several clouds at
different physical conditions are taken into account.

\section {Conclusions}

The ISO spectra of the coronal lines [O IV] 25.9 $\mu$, [Ne V] 14.3 $\mu$,
[Mg VIII] 3.02 $\mu$ and [Si IX] 2.58 $\mu$
for a sample of bright Seyfert type 2 and 1 galaxies are presented.  To 
produce   the ionization states that  give  rise to those
fine-structure lines,  energies in the 50 - 300 eV
range are required. Accordingly, by studying those lines,  
 clues on that soft part of the 
ionization continuum can be  derived.

The emission lines 
[O IV] 25.9 $\mu$ and [Ne V] 14.3 $\mu$ are  found to be common strong
features in the Seyfert spectra regardless of their type.
On the other hand, 
 [Mg VIII] 3.02 $\mu$ and [Si IX] 2.58 $\mu$ are much weaker and
for most cases, just an upper limit is provided.

The ratio of  these coronal lines
relative to [O IV] 25.9 $\mu$  are compared to
model predictions derived under pure photoionization conditions.
Two type of UV 
ionization continuum, namely,  a power-law and 
a hot blackbody, are used. 
Within a reasonable range of values for the ionization parameter U,
and the optical depth of the clods, $\tau_{ly}$, both type of models
appear  indistinguishable on   the basis of these data alone.

However,  some constrains on  the ionization parameter
and the optical depth of the emitting clouds  can be derived.
  Independent on the  UV continuum model selected,  values of the
ionization parameter below 0.1
are required. This is  mostly  to account for the systematic 
low [SiIX]/[OIV] ratios
  indicated by  the data. The  optical depth of the emitting clouds,
$\tau_{ly}$,  is more UV-continuum model dependent 
 but in general  values about 1 or larger
are indicated for most of the sources. Specifically, the range of
$\tau_{ly}$ indicated by the data  varies from
 optically thin to optically thick clouds
in the case of  an UV power law distribution whereas
optically thick clouds are mostly indicated in the case of an UV blackbody.

Focusing on the brightest Seyfert in the sample,
 Circinus, NGC 1068 and  NGC 4151, for which accurate measurement of
 the above coronal lines is available, a more clear pattern arises.
They all show a  decreasing
 trend  of the ratio of  the coronal lines
relative to [O IV] 25.9 $\mu$ as a function of the ionization
potential (required to produce the emitting ion).
Our best approximation to that pattern is obtained by assuming an UV 
black-body with a cutoff energy in the 100 - 120 eV range. 
Furthermore, U values
 below 0.01 and  $\tau_{ly}$ larger than 1 are required. In fact, 
 low U values couple to large values of $\tau_{ly}$ favor the  
presence of lower ionization ions such as  O$^{+3}$ compared to 
Si$^{+8}$, consequently decreasing the  [SiIX]/[OIV] ratio.
In addition, if the cloud is optically thick, the volume emitting
[O IV] is larger, and the  [SiIX]/[OIV] ratio is still lower as 
seen in NGC 4151.

Other facts however are expected to contribute as well to that pattern. 
First the all,  these line ratios depend on the
abundance ratios like Ne/O, Mg/O and Si/O whereas 
 simply solar ratios are  assumed in our models. Second, 
a reliable  determination of the ionizing 
radiation spectrum could only be obtained if
a mixing of clouds with different U and $\tau_{ly}$
is chosen to fit several emission-line intensities
in a more complete multiwavelength study. \\[2mm]

{\bf Acknowledgments: }
It is a pleasure to thank  Dietmar Kunze and Eckhard Sturm 
from the ISO Spectrometer  Data Center at MPE
for their constant support in the processing of the SWS data. The
referee, Dieter Lutz, is acknowledged for his critical review. 
MAP is thankful to Fapesp for allowing her visit to IAGUSP, Brazil. 
This work is partially supported by FAPESP, CNPq and
PRONEX/Finep, Brazil.

\def\reference {\par \noindent \parshape=6 0cm 12.5cm 
0.5cm 12.5cm 0.5cm 12.5cm 0.5cm 12.5cm 0.5cm 12.5cm 0.5cm 12.5cm}


\reference Binette, L., 1998, MNRAS 294, L47
 
\reference Binette, L., Wilson, A, Raga, A. \& 
Storchi-Bergmann, T. 1997 A\&A 327,909

\reference Contini, M., Prieto, M. A., \& Viegas, S. M.  1998a,
ApJ 492, 511

\reference Contini, M., Prieto, M. A., \& Viegas, S. M.  1998b,
ApJ 505, 621

\reference Ferland et al. 1995, in The Analysisof Emission Lines,
STScI Symp. Ser. 8, ed. R. E. Williams \& M. Livio (Cambridge:
Cambridge Univ. Press), 143

\reference Feuchtgruber et al 1997, ApJ 487, 962

\reference Genzel, R. et al, 1998, ApJ 498, 579

\reference Gruenwald, R. \& Viegas, S. M.  1992, ApJS 78, 153

\reference Lutz, D., Sturm. E., Genzel, R. Moorwood, A. \& Sternberg, A. 1997, 
Ap\&SS 248, 217 

\reference Lutz, D., Kunze, D., Spoon H., Thornley, M. 1998, A\&A 333, L75 
 
\reference Moorwood, A.F.M., Lutz, D., Oliva, E., Marconi, A.
Netzer, H., Genzel, R., Sturm, E., \& de Graauw, Th. 1996, A\&A 315, L109

\reference Penston et al, 1984, MNRAS 208, 347

\reference Schaeidt, S. G. et al., 1996, A\&A 315, L55

\reference  Sturm, E., Alexander, T.,\, Lutz, D., T., Sternberg, A., 
Netzer, H. \& Genzel, R., 1999, ApJ 512, 197


\newpage

\topmargin 0.01cm
\oddsidemargin 0.01cm
\evensidemargin 0.01cm

\begin{table}
\begin{center}
\centerline{Table 1: ISO coronal line fluxes for Seyfert galaxies}
\begin{tabular}{ l l l l l l l }\\ \hline   \hline
Name& Syf& V &    [SiIX]&    [MgVIII]&      [NeV] &   [OIV] \\ \hline 
  & Type  & Km/s   & 2.59$\mu$ & 3.03$\mu$  &14.3$\mu$  & 25.9$\mu$  \\ \hline 
 
N5548 & 1 &5152 & $<$4.9e-21 & $<$2.e-21 &  8.9e-21 $\pm30$ & 7.5e-21 $\pm10$ \\ 
N5929 & 2 &2490  &$<$ 4.9e-21   & $<$ 1.1e-21  & $<$3.5e-21 $\pm50$ & 
2.1e-21 $\pm40$ \\     
Mrk817 &1 &9436  &  3.1e-21 ?    & 1.1e-21 ? & 6.9E-21 $\pm30$ & 
3.12e-21 $\pm13$ \\ 
Mrk335 &1 &7688 &$<$ 1.e-21  &$<$ 1.e-21 &$<$  1.2e-20 & 1.9e-20 $\pm30$ \\    
Mrk266 &2 &8360 &  $<$ 5.8e-21&   1.8e-21 ?  &   5.e-21 $\pm30$ &  
2.1e-20 \\ 
Mrk533 &2 &8670 & $<$  7.2e-21 & 1.8e-21 ? & 1.1e-20 $\pm20$ & 
3.2e-20 \\ 
Mrk334 &2 &6582 &  $<$ 2.9e-21 &$<$ 1.3e-21     &  2.7e-21 $\pm30$ & 
4.3e-21  \\  
N1144 &2 &8648 &$<$   4.9e-21  &  $<$ 1.4e-21 & 7e-21 $\pm30$ & 8e-21    
\\  
N5033 &1 & 876 & $<$  3.57e-21 & $<$ 1e-21     &   6.1e-21 $\pm20$& 
1.2e-20\\  
CIRCI &2 & 436 &   1.8e-20 &   6.5e-20   &     44e-20&   72e-20\\    
N1068 &2 &1140 &   4.1e-20 &   14e-20   &    97e-20 &  160e-20\\      
N4151 &1.5 & 980 &  2.3e-21 $\pm20$&   6.2e-21  &        55e-21&   20e-20
 \\ 
\hline \hline
\end{tabular}
\end{center}

Fluxes are in $W~cm^{-2}$. Upper limits are derived from the most
promising feature in the spectra; otherways a 3 sigma limit is given. 
Errors  in  the flux (indicated in percent)
 only reflect  the uncertainty in defining the
continuum level; when no indicated, the errors are within 10\%. 
Uncertain identifications are marked with a question mark.
Data from Circinus are taken form Moorwood et al (1996);  NGC 1068 from
Fig. 2 in Lutz et 
al (1997); NGC 4151 from Sturm et al. (1999).
Redshift and Seyfert type are taken from NED.
\end{table}

\clearpage
{\bf Figure Captions}\\

Figure 1a,b - ISO spectra for the sample of observed Seyfert 1 (a) and 2
 (b) galaxies. 
Each row corresponds to a given galaxy  identified in the first panel.
Y-axis represents absolute flux in Jy; X-axis is  the redshifted spectrum
measured with respect to the position of the expected line in
 $\mu$. The given range is equivalent to about $\pm$ 1000 km/s about the
 central position. 
First panel in each row corresponds to [MgVIII 3.02 $\mu$]; second to 
[SiIX]2.58 $\mu$, third to 
[NeV] 14.3 $\mu$; forth shows [OIV]25.9 $\mu$ and [FeII] 25.98 $\mu$
(in this case, the x-axis  is measured 
 with respect to the position of the [OIV] line, the total range is
 about 3000 km/s). 

Figure 2a,b - Coronal emission-line intensities relative to
[O IV] 25.9 $\mu$ as a function of the the ionization
energy required to produce the emitting ion
for, (a) Seyfert 1 galaxies: NGC 5548(solid triangle), Mrk 817(solid square),
Mrk 335(solid circle), NGC 5033(star), NGC 4151(circle),
and (b) Seyfert 2 galaxies: NGC 5929(solid triangle), Mrk 266(solid square),
 Mrk 533 (solid circle), Mrk 334(star),  NGC 1144(circle),
Circinus(empty square), NGC 1068 (empty triangle)
Data for Circinus, NGC 1068 and NGC 4151 are joined by lines.

Figure 3a -  Coronal emission-line intensities, relative to
[O IV] 25.9 $\mu$, from photoionization models with 
a blackbody UV ionizing radiation. The top panel show the results
for the models with U = 0.2 and $\tau_{ly}$ = 0.2, 1., 2., 3., 5.
and 30 (solid, dotted, dashed, long-dashed, dot-dashed and
solid, respectively). A model corresponding to U = 0.006 and
$\tau_{ly}$ = 30 is also showed (bottom dashed line).
 The bottom panel show 
optically thick models with different values of the ionizing
parameter U = 0.4 (higher solid line) to 0.006 (bottom dashed line).

Figure 3b -  Coronal emission-line intensities, relative to
[O IV] 25.9 $\mu$, from photoionization models with 
a power-law  UV ionizing radiation. The top panel show the results
for the models with U = 0.2 and $\tau_{ly}$ = 0.2, 1., 2.,, 5.
and 30 (solid, dot-dashed, dotted, dashed,  and
solid, respectively). A model corresponding to U = 0.006 and
$\tau_{ly}$ = 30 is also showed (lower dashed line). 
The bottom panel show 
optically thick models with different values of the ionizing
parameter U = 0.4 (higher solid line) to 0.006 (lower solid line).

Figure 4 - Diagnostic diagrams for the IR coronal lines: data
The triangles correspond to Syf 1  and solid circles to Syf 2.
Those galaxies with uncertain  detection of  [MgVIII] (Mrk 266 and Mrk
533) and [SiIX] (Mrk 817) are identified. Also the brightest Seyfert
NGC 1068, NGC 4151 and Circinus are marked.

Figure 5a,b - Diagnostic diagrams for the IR coronal lines: data plus models.
Theoretical results correspond to models with UV blackbody
radiation and Ec=100 eV. Each curve correspond to models  with
different values of U
and a given $\tau_{ly}$ = 0.2, 1.,   5. and 30 (solid, dot-dashed, 
dashed,  and  solid, respectively).
The observed data are overimposed:
the triangles correspond to Sy 1  and solid circles to Sy 2.
The thick line correspond to a model with cut-off energy Ec= 120 eV
and  $\tau_{ly}$ = 1.5.

Figure 6a,b - Diagnostic diagrams for the IR coronal lines.
Theoretical results correspond to models with UV power-law
radiation. Each curve correspond to models  with different values of U
and a given $\tau_{ly}$ = 0.2, 0.3, 1. and 30
(solid, short-long dash, dot-dashed, and 
solid, respectively).
The observed data are overimposed:
the triangles correspond to Sy 1  and solid circles to Sy 2 results.

Figure 7 - Coronal emission-line intensities relative to
[O IV] 25.9 $\mu$ as a function of the the ionization
energy required to produce the emitting ion
for NGC 1068 (square), Circinus (triangle) 
and NGC 4151 (circle). Two set of models for an UV
black-body radiation and cut-off energy Ec= 120 eV
 are displayed: dot lines refer to U=0.01 and $\tau_{ly}$= 1.5 and
30; dash lines refer to  U=0.005 and same $\tau_{ly}$ as before.

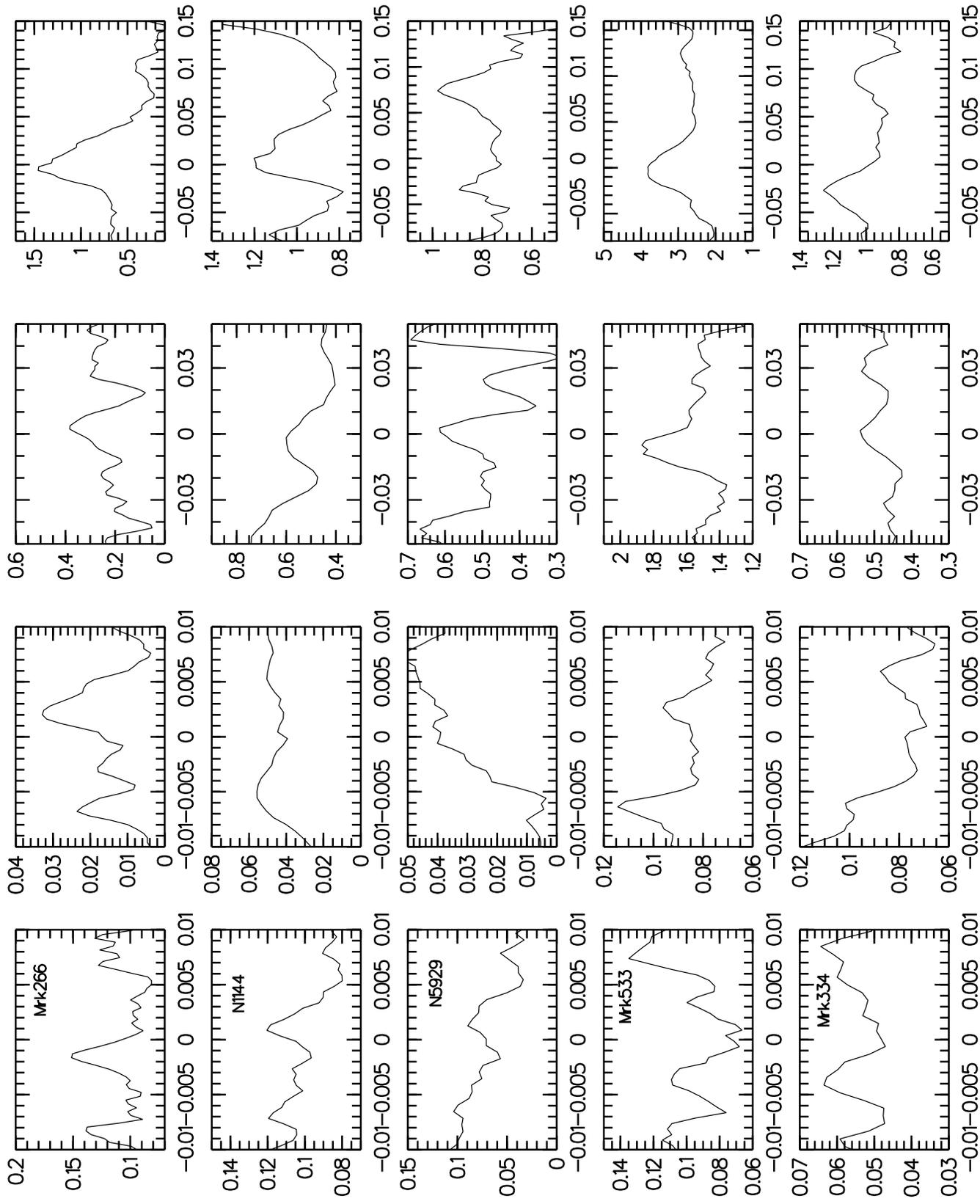
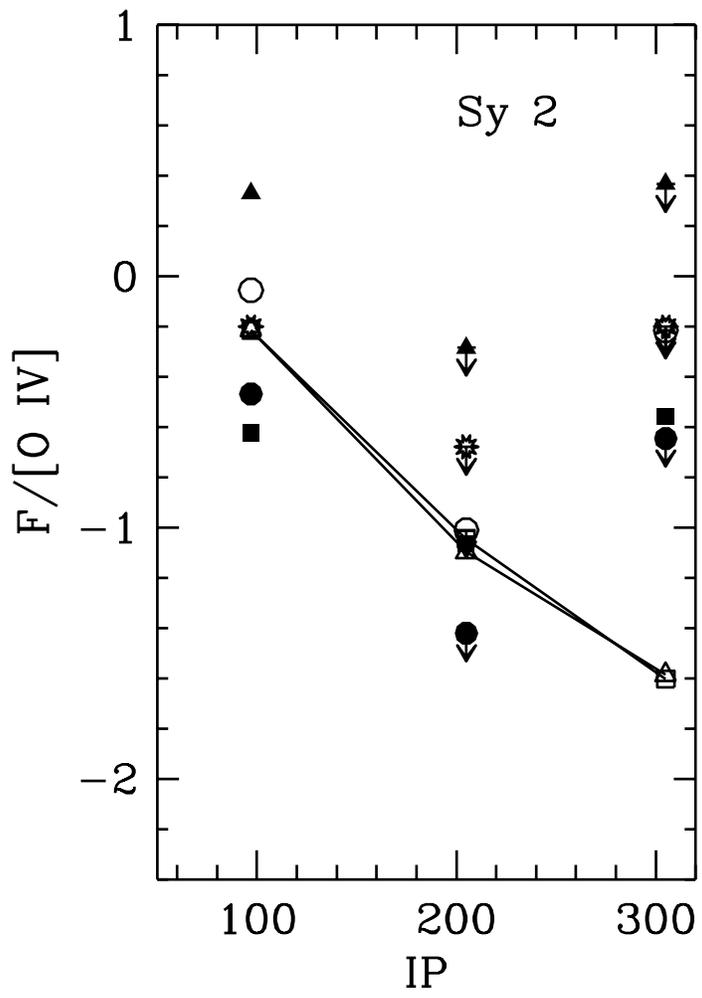
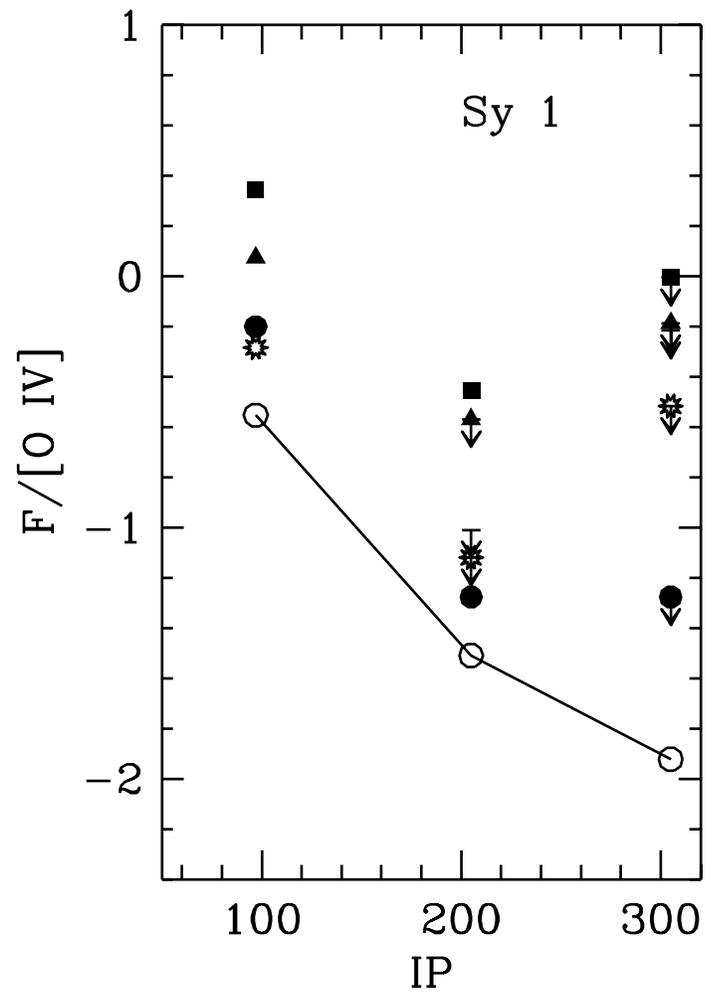
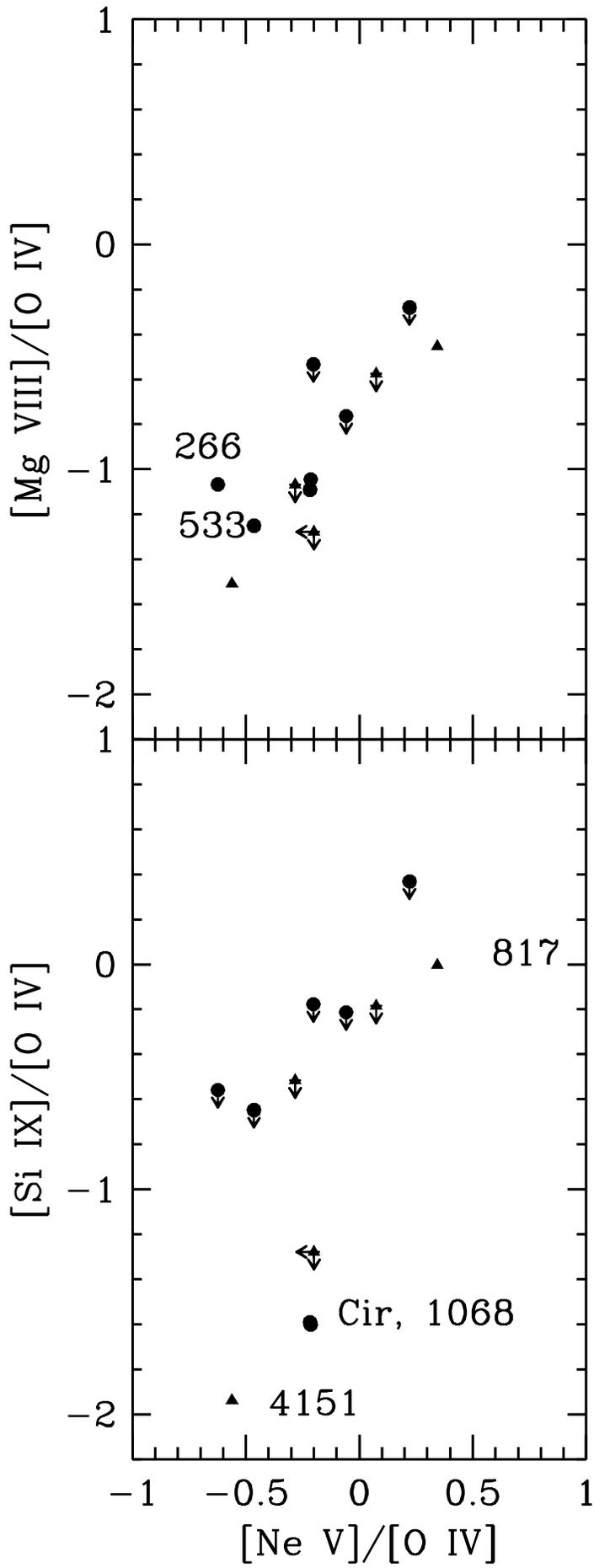
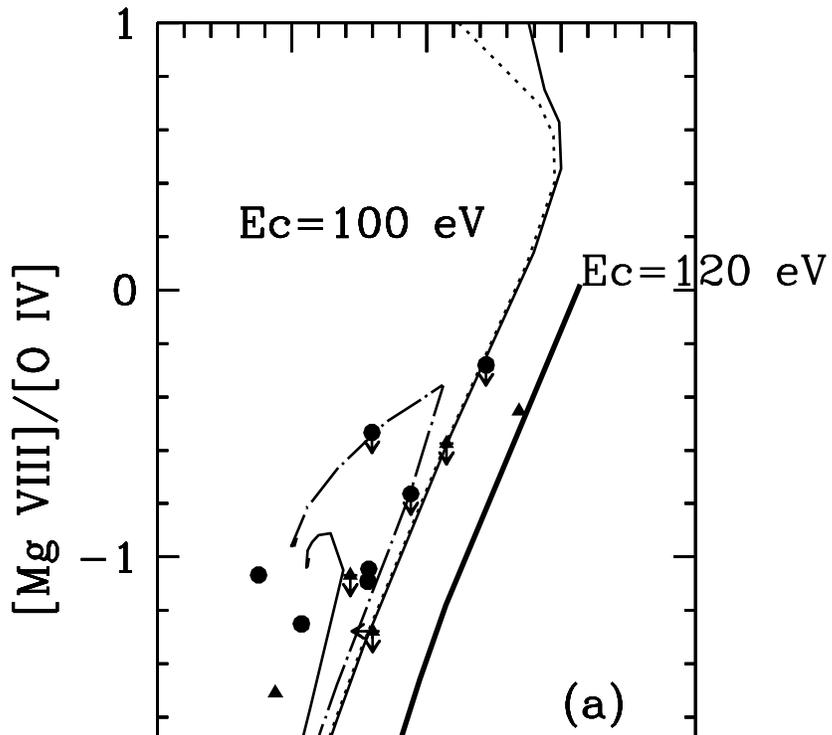
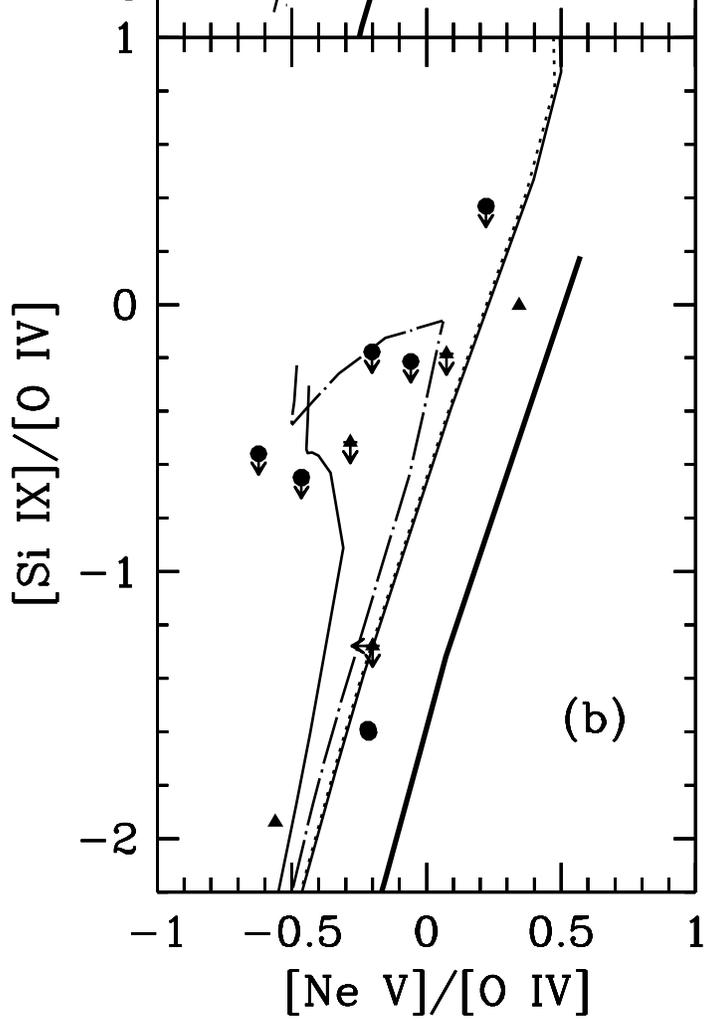
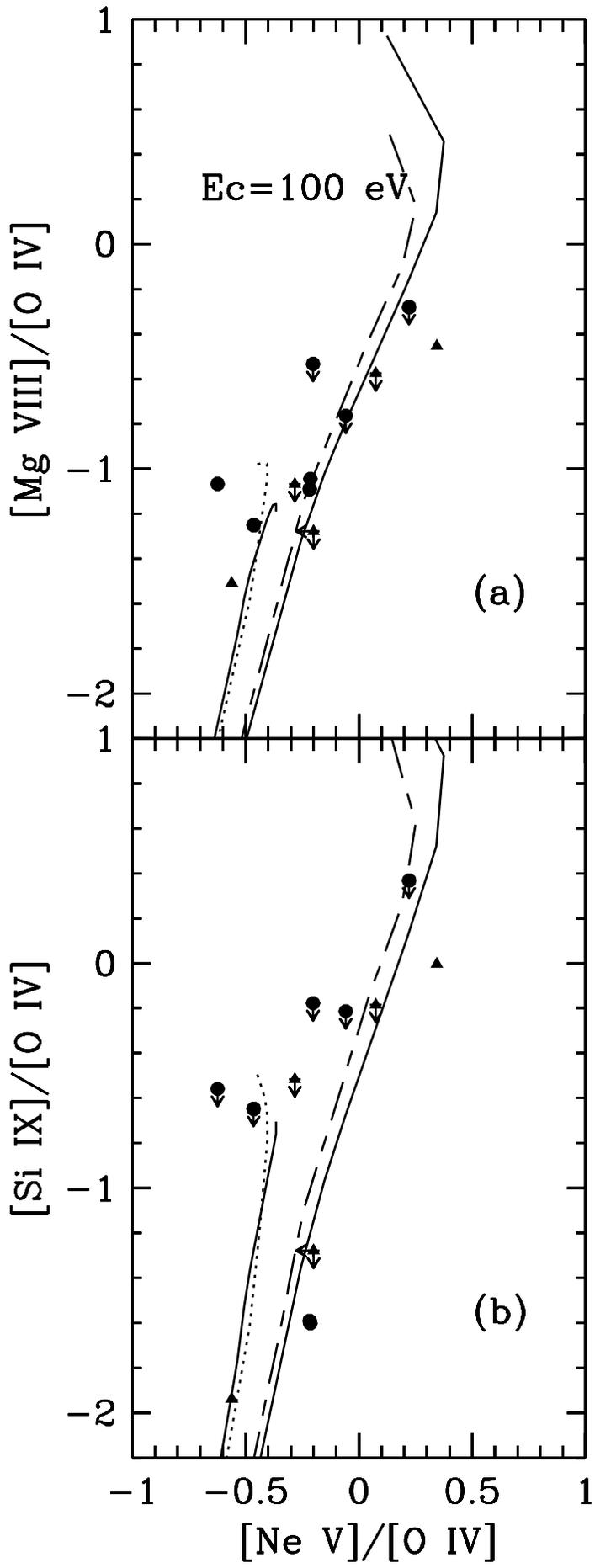
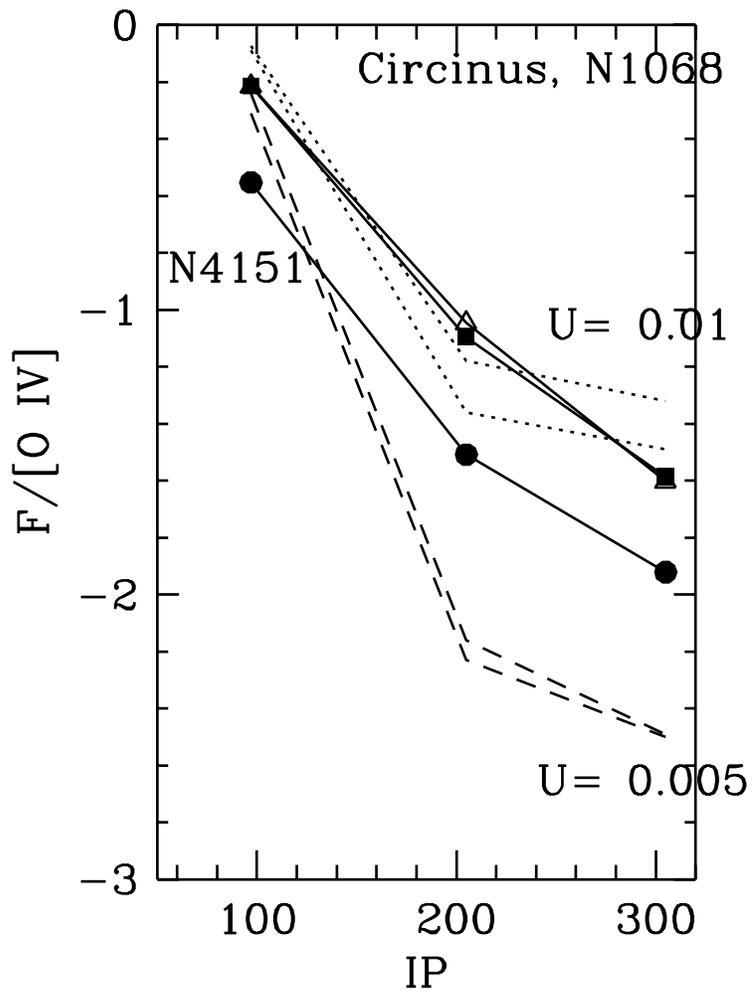
\end{document}